\documentclass[conference]{IEEEtran}
\IEEEoverridecommandlockouts

\usepackage{amsmath,amssymb,amsfonts}
\usepackage{algorithm}
\usepackage{algorithmicx}
\usepackage{algpseudocode}
\usepackage{setspace}
\usepackage{bm}
\usepackage{graphicx}
\usepackage{textcomp}
\usepackage{xcolor}
\usepackage{tikz}
\usetikzlibrary{quantikz2}
\usepackage{float}
\usepackage{subcaption}
\usepackage[T1]{fontenc}
\usepackage{verbatim}
\usepackage{hyperref}
\usepackage{makecell}
\usepackage{booktabs}
\usepackage{array} 
\usepackage{amsthm}
\usepackage{graphbox} 
\usepackage{adjustbox}
\usepackage{balance}
\usepackage{multirow}
\usepackage[capitalize]{cleveref}
\usepackage{footnote}
\usepackage{bm}
\usepackage{threeparttable}
\usepackage{caption}
\usepackage{lipsum}

\def\BibTeX{{\rm B\kern-.05em{\sc i\kern-.025em b}\kern-.08em
    T\kern-.1667em\lower.7ex\hbox{E}\kern-.125emX}}

\theoremstyle{remark}

\theoremstyle{definition}

\begin{document}

\title{Quantum Paths: a Quantum Walk approach\\
\thanks{This work has been funded by the European Union under Horizon Europe ERC-CoG grant QNattyNet, n.101169850. Views and opinions expressed are however those of the author(s) only and do not necessarily reflect those of the European Union or the European Research Council Executive Agency. Neither the European Union nor the granting authority can be held responsible for them.}
}

\author{Claudio Pellitteri,  Marcello~Caleffi,~\IEEEmembership{Senior~Member,~IEEE}, Angela~Sara~Cacciapuoti,~\IEEEmembership{Senior~Member,~IEEE}
    \thanks{This work has been accepted in Proceedings of IEEE International Conference on Quantum Computing and Engineering 2025.}
    \thanks{The authors are with the \href{www.quantuminternet.it}{www.QuantumInternet.it} research group, University of Naples Federico II, Naples, 80125 Italy.}}

\maketitle

\begin{abstract}
The quantum switch, a process enabling a coherent superposition of different orders of quantum channels, has garnered significant attention due to its ability to enable noiseless communications  through noisy channels, such as entanglement-breaking channels. However, its practical implementation and scalability remain challenging. In contrast, the spatial superposition of quantum channels is more accessible experimentally and has been shown to enhance channel capacity, although it does not match the performance of the quantum switch. In this work, we present preliminary theoretical results demonstrating that, by applying tools of the quantum random walk framework to the spatial superposition of channels, it is possible to replicate the output of a quantum switch. These findings suggest a promising and more feasible route to emulate the quantum switch, offering both practical advantages and interpretative clarity.
\end{abstract}

\begin{IEEEkeywords}
Quantum Switch, Quantum Paths, Quantum Walk, Quantum Internet, Quantum Communications, ERC-CoG QNattyNet.
\end{IEEEkeywords}

\section{Introduction}
\label{sec:1}
In quantum networks, it is possible to realize genuine-quantum instantiations of quantum communication channels, by exploiting the peculiarity of quantum carriers to propagate thought multiple space-time trajectories. This has no-counterpart in classical networks and it gives rise to the concept of \textit{quantum paths} \cite{CalSimCac2023}.

The quantum Switch realizes a particular form of \textit{quantum path} via the quantum superposition of orders of different quantum communication channels. The quantum Switch has gained a lot of attention and interest in recent years due to its peculiarities \cite{ChiDarPer2013,EblSalChi2018,GosGiaRom2018}. Indeed, it has been shown that through a quantum switch, it's possible to enable noiseless communications through noisy channels, such as entanglement breaking channels \cite{CalCac2020,CalSimCac2023}. Through a quantum switch it is also possible to engineer universal quantum gates \cite{SimCalIlRomCac2024}. Although photonic implementations of a quantum switch have been realized in controlled environments \cite{Guo2020,RubRozFeixAraZeuProcBruWal,Stromberg2023,Pro-19,ProMoqAra-15}, its real-world implementation and scalability remain challenging \cite{EscMon2023}. Furthermore, a scientific debate about the interpretation of the results of the aforementioned experiments is still open. 

Thankfully, a different type of \textit{quantum path}, relatively easy to implement, exploits a spatial superposition of quantum channels \cite{Pang2023,ChiKri-18}. This form of quantum path has been proven to have a higher quantum channel capacity than the one achievable by exploiting classical propagation \cite{ChiKri-18}. However, the quantum switch performance overcomes the ones of the spatial superposition. 

In this paper, differently from the state-of-the-art, we prove that it is possible to obtain the same performance of a quantum switch, by exploiting the spatial superposition of quantum channels. For doing this, we hybridize the quantum path setup in its spatial superposition form, with tools coming from the quantum random walk framework, namely, the quantum counterpart of the classical random walk. 

More into details, the quantum random walk framework describes the evolution of the position of a quantum particle in a lattice, conditioned to the measurement of the quantum state of a controlled qubit, referred to as quantum coin. In every step, the quantum coin is subjected to a unitary rotation, the quantum-analogue of tossing a coin. This framework has already been demonstrated to be successful for engineering  unitary channels and for solving, in polynomial time, problems that would take exponential time to be solved classically \cite{Kempe2005}.
\begin{figure}[t]
    \centering
\includegraphics[width=1.\columnwidth]{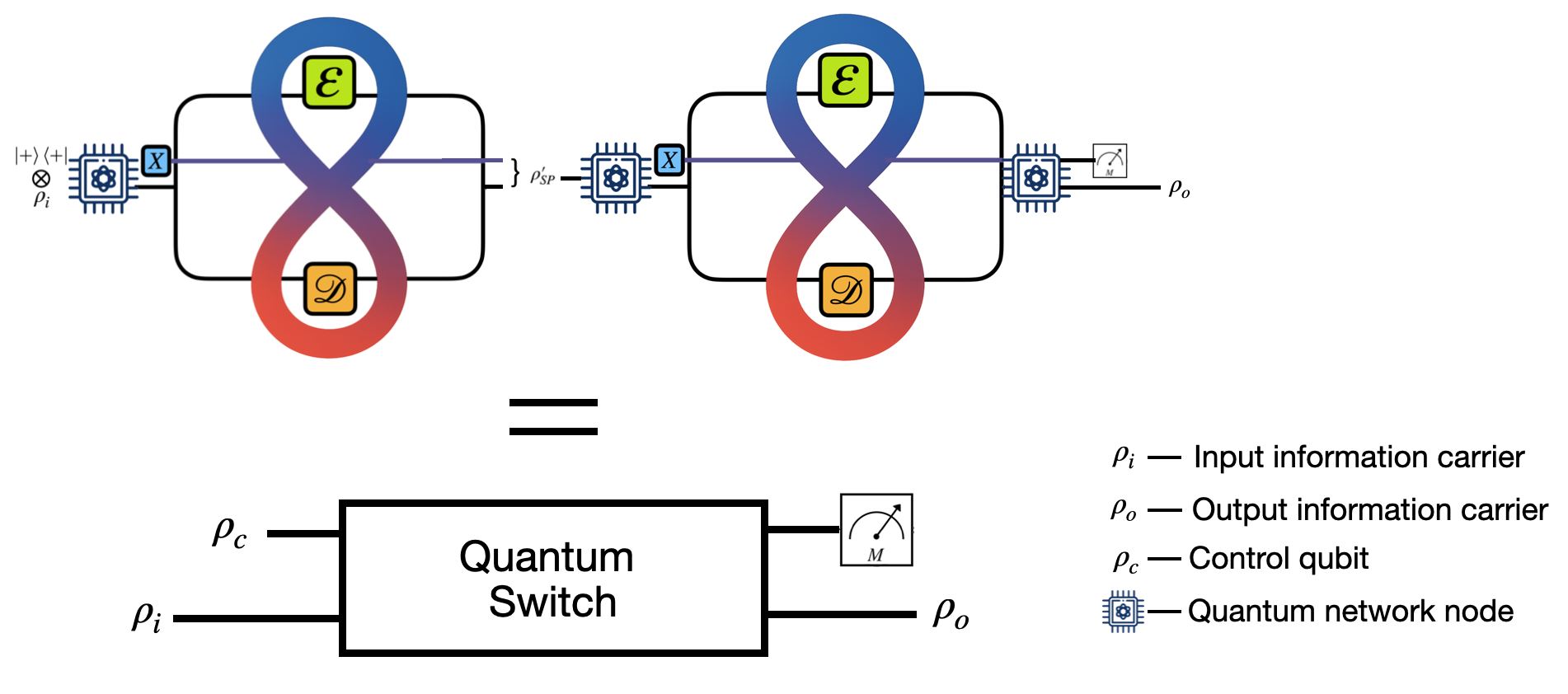}
\caption{Pictorial representation of the main result of this work: mimicking the quantum switch behavior via spatial superposition. Specifically, the network nodes are connected via two-hops. The evolution of the state of the information carrier through the two-hops spatial superposition of the quantum channels $\mathcal{E}$ and $\mathcal{D}$ is equivalent to the output of the quantum switch, exploiting the causal-order superposition of the considered channels.}
\label{fig: 1}
\end{figure}

In this work, by leveraging the quantum walk, we prove some theoretical results, which shed the light on the possibility to recover the input-output relationship of the quantum switch by applying tools proper of the quantum walk to the channel spatial superposition, as pictorially depicted in \cref{fig: 1}. These results, although preliminary, are quite remarkable since they show how it's possible to implement in an easier way the quantum switch and to give, at least at a simulation level, a clear interpretation of how it arises. 

\section{The two types of Quantum Path}
\label{sec:2} 
\subsection{Spatial Superposition of quantum channels}
\label{sec: 2.1}
\begin{figure}[t]
    \centering
\includegraphics[width=1.\columnwidth]{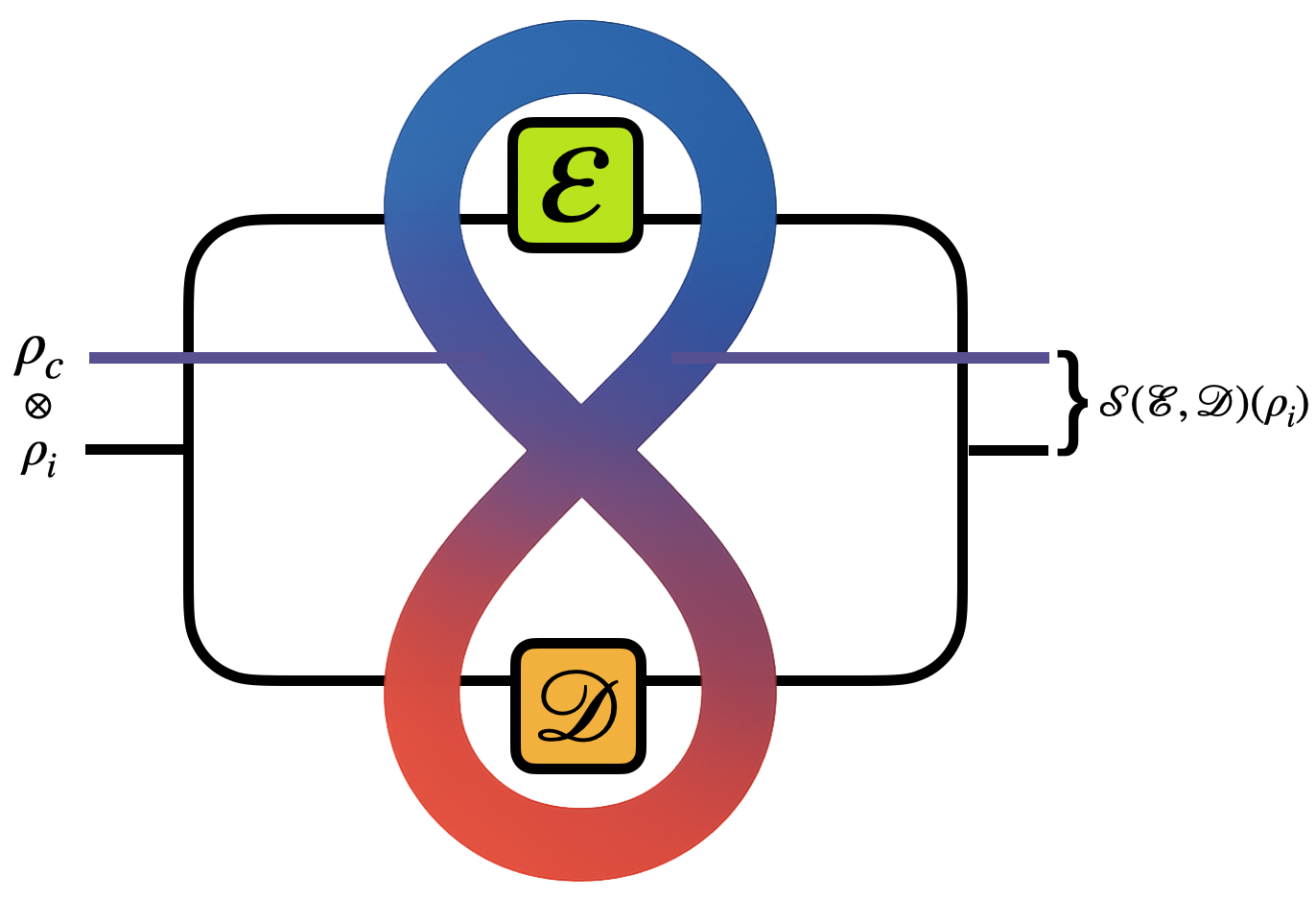}
\caption{Pictorial representation of the spatial superposition of quantum channels. Based on the state of the control qubit, the quantum carrier can go either through one of the channels ($\mathcal{E}$ or $\mathcal{D}$) or through a quantum superposition of the two channels. Specifically, if the control qubit is initialized in $\ket{0}\bra{0}$ ($\ket{1}\bra{1}$), the information carrier propagates through channel $\mathcal{E}$ (channel $\mathcal{D}$), but if the control qubit is prepared in $\ket{+}\bra{+}$ ($\ket{-}\bra{-}$) then the information carrier propagates through a spatial superposition of both the channels.}
\label{fig: 2}
\end{figure}

In order to describe the spatial superposition of two channels, Chiribella et al. developed a formalism \cite{ChiKri-18} that augments the dimension of the associated Hilbert space to take into account the possibility of the channel not being used, via the vacuum extension of the quantum channels. This extension of the quantum channels, by introducing the vacuum state $\ket{vac}$, allows to describe the state evolution of an information carrier that goes either through one of the two channels or through a quantum superposition of the two channels. A pictorial representation of this type of quantum path via spatial superposition of quantum channels is given in \cref{fig: 2}.
Formally, the spatial superposition of the channels is described by the map:
\begin{equation}
\label{eq:01}
    \mathcal{S}(\mathcal{E},\mathcal{D})=\sum_{ij}S_{ij}\left(\rho\otimes\rho_c\right) S^\dagger_{ij},
\end{equation}
where $S_{ij}$ is given as follows:
\begin{equation}
  \label{eq:02}
S_{ij}=E_i\beta_j\otimes\ket{0}\bra{0}+\alpha_i D_j\otimes\ket{1}\bra{1}.  
\end{equation}
In \eqref{eq:02}, $\{E_i\}$ and $\{D_i\}$ are the Kraus operators of the considered quantum channels and $\alpha_i$ ($\beta_i$) are the vacuum amplitudes relative to  $E_i$ ($D_i$). These amplitudes are normalized, i.e.,  $\sum_i|\alpha_i|^2=1$ and are not unique. In fact their values depend strongly on the type of interaction between the system and the environment.
Accordingly, the states of the information carrier and the control qubit -- which takes into account which path is traversed -- evolve as:
\begin{align}
\label{eq: 3}
\rho_{SP}&=\mathcal{S}(\mathcal{E},\mathcal{D})(\rho_i\otimes\rho_c)=\sum_{ij}S_{ij}(\rho_i\otimes\rho_c) S^\dagger_{ij}=\\\nonumber
&=\sum_{ij}(E_i\beta_j\otimes\ket{0}\bra{0}+D_j\alpha_i\otimes\ket{1}\bra{1})(\rho_i\otimes\\\nonumber
&\otimes\rho_c) (E^\dagger_i \beta^*_j\otimes\ket{0}\bra{0}+D^\dagger_j\alpha^*_i\otimes\ket{1}\bra{1}).
\end{align}

\subsection{Causal-order Superposition of quantum channels}
\begin{figure}[t]
    \centering
\includegraphics[width=1.\columnwidth]{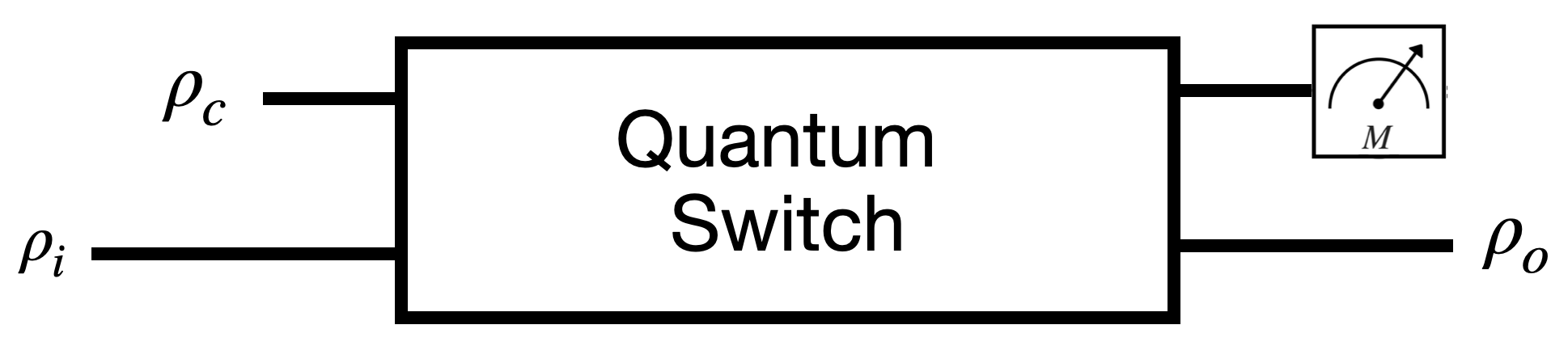}
\caption{Pictorial representation of the quantum switch. The information carrier is sent through a superposition of different causal orders. Based on the quantum state of the control qubit, the informational carrier can go through an ordered sequence of channels $\mathcal{E}\rightarrow\mathcal{D}$ (equivalently $\mathcal{D}\rightarrow\mathcal{E}$) or a quantum superposition of the two orders.}
\label{fig: 3}
\end{figure}

As aforementioned, another type of quantum path is the one realized via the causal-order superposition of quantum channels, as represented in \cref{fig: 3}, through the so-called quantum switch.
Formally, the quantum switch is described by the action of a supermap $\mathcal{S}$, whose Kraus operators are given by:
\begin{equation}
\label{eq:04}
    S_{ij}=E_i D_j\ket{0}\bra{0}+ D_jE_i\ket{1}\bra{1},
\end{equation}
where $\{E_i\}$ and $\{D_i\}$ are the Kraus operators of the considered channels, $\mathcal{E}$ and $\mathcal{D}$.
As mentioned in \cref{sec:1}, it has been proved that the quantum switch can herald noiseless communications, namely, quantum channel capacity equals to 1, also in extreme cases such as the presence of entanglement breaking channels. This marvelous property makes the quantum switch key in quantum networks, for entanglement distribution and quantum information transmission. In particular in \cite{CalSimCac2023,CalCac2020} it has been shown how the quantum switch makes possible to send quantum information even when the quantum capacity of one of the two quantum channels is zero, by overcoming so the bottleneck inequality. Although all these remarkable properties, there still exist technical difficulties in the experimental implementation of the quantum switch as well as in its scalability \cite{EscMon2023}.


\section{Mimicking the quantum Switch behavior via Spatial
Superposition}
\label{sec:3}
\subsection{Quantum Walk}
The extension to the quantum realm of the random walk process has attracted the attention of both physicists and engineers for its impact on quantum information and computation \cite{Chan2010}. Here, for the sake of clarity, we briefly summarize the one-dimensional discrete time quantum random walk (DTQW) properties. 

The classical unidimensional random walk can be seen as the evolution of a particle that moves to the right or to the left on a grid based on the result of a binary probabilistic event, e.g. the tossing of a coin. In order to extend this kind of dynamics in the quantum domain, it must be considered a N-dimension Hilbert space $H_p$ spanned by the states {$\ket{i}$, i=1,...,N}, which describes the position of a particle (quantum walker) in a one-dimensional gridline and a two dimension Hilbert space spanned by the states {$\ket{0}$,$\ket{1}$} which play the role of the coin. The dynamics of a quantum particle is described by the action of the following unitary: 
\begin{equation}
    \label{eq: 5}
    U=T(I\otimes C),
\end{equation}
where $C$ is some unitary rotation of the coin state, $I$ is the N-dimensional identity and $T$ is the conditional shift operator
\begin{equation}
    \label{eq: 6}
    T=\sum_i \ket{i+1}\bra{i}\otimes\ket{0}\bra{0}+\sum_i \ket{i-1}\bra{i}\otimes\ket{1}\bra{1}.
\end{equation}
The most evident difference with its classical counterpart is that the average displacement of a particle in a DTQW is not null \cite{Kempe2003}. Moreover, by choosing a balanced coin operator and assuming that the coin state is in a quantum superposition of its two basis state, the displacement probability is symmetric, since  the walker is experiencing some sort of superposition of paths on the gridline.

\subsection{Combining Quantum Walk with Spatial Superposition of channels}
\label{sec:4}
Stemming from the previous section, we infer similarities between the conditional shift in the DTQW and the supermap describing spatial superposition of channels. Indeed, both apply different evolutions on the information carrier (the walker in the quantum walk framework) accordingly to the state of the control qubit, which describes the traversed path. 

In the following we present the proposed hybridization of the quantum walk framework and the quantum path setup in the form of spatial quantum superposition for recovering the output of a quantum switch. Specifically, we consider a two-hops evolution of an information carrier as represented in Fig.~\ref{fig: 1}, where each step is described by the quantum channel:
\begin{equation}
    \label{eq: 7}
\mathcal{W}=\mathcal{S}\circ(\mathcal{I}\otimes\boldsymbol{C})
\end{equation}
where $\mathcal{S}$ is the superposition channel described in \cref{sec:2}, $\mathcal{I}$ is the identity matrix acting on the system and $\mathcal{C}$ is the coin operator which acts on the control qubit. We now show that it is possible to obtain the quantum switch input-output relationship via a two-hops evolution via the channel in \cref{eq: 7}. 

\subsubsection{Unitary channels}
\label{sec:2.1}
We study the case in which the quantum carrier experiences a two-hops evolution in the case in which the two quantum paths are respectively described by the unitary channels $U_1$ and $U_2$. The superposition of two unitary channels is described by the unitary map:
\begin{equation}
\label{eq: 8}
S=U_1\otimes\ket{0}\bra{0}+U_2\otimes\ket{1}\bra{1}.
\end{equation}

Let us assume that the information carrier is prepared in a generic state $\rho_i$ and that the control qubit is initially in the state $\ket{+}=1/\sqrt{2}(\ket{0}+\ket{1})$. And we set the coin operator as a Pauli $\boldsymbol{X}$ gate. After the first hop, the state of both the information carrier and the control qubit is:
\begin{align}
\label{eq:9}
\rho_{SP}'=&S(\rho_i\otimes\ket{+}\bra{+})S^\dagger=\\\nonumber
&\frac{1}{2}(U_1\rho_i U^\dagger_1\otimes\ket{0}\bra{0}+U_1\rho_i U^\dagger_2\otimes\ket{0}\bra{1}+\\\nonumber
&+U_2\rho_i U^\dagger_1\otimes\ket{1}\bra{0}+U_2\rho_i U^\dagger_2\otimes\ket{1}\bra{1}).
\end{align}
By applying again the $\mathcal{W}$ channel, we obtain: 
\begin{align}
    \label{eq: 10}
    \rho''_{SP}=&S((\mathcal{I}\otimes X)\rho'_{SP}(\mathcal{I}\otimes X))S^\dagger=\\\nonumber
&\frac{1}{2}(U_1 U_2\rho_i U^\dagger_2U^\dagger_1\otimes\ket{0}\bra{0}+U_1U_2\rho_i U^\dagger_1U^\dagger_2\otimes\ket{0}\bra{1}+\\\nonumber
&+U_2U_1\rho_i U^\dagger_2U^\dagger_1\otimes\ket{1}\bra{0}+U_2U_1\rho_i U^\dagger_1U^\dagger_2\otimes\ket{1}\bra{1}).
\end{align}
It's easy to see that \cref{eq: 10} is exactly the outcome of a quantum switch that describes the superposition of causal orders of the two considered unitary channels.

\subsubsection{Non-unitary Channels}
It's possible to generalize the above analysis to two arbitrary channels $\mathcal{E}$ and $\mathcal{D}$. Specifically, by considering the same states and coin operator used in the previous sub-section, after the first hop, the state of carrier and control qubit is given by:
\begin{align}
    \label{eq: 11}
    \rho'_{SE}=&\sum_{lj}(|\beta_j|^2 E_l\rho_i E^\dagger_l\otimes\ket{0}\bra{0}+\alpha_l^*\beta_j E_l\rho_i D^\dagger_j\otimes\ket{0}\bra{1}+\\
    &+\alpha_l\beta^*_jD_j\rho_i E^\dagger_l\otimes\ket{1}\bra{0}+|\alpha_l|^2D_j\rho_i D^\dagger_j\otimes\ket{1}\bra{1},\nonumber
\end{align}
which is just the spatial superposition of the two channels, due to the fact that the control qubit is initially prepared into an eigenstate of the coin operator. By applying again the $\mathcal{W}$ channel, one obtains: 
\begin{align}
   \label{eq: 12}
    \rho''_{SE}=&\sum_{sjlm}(|\alpha_s||\beta_m|^2 E_l D_j\rho_i D^\dagger_j E^\dagger_l\otimes\ket{0}\bra{0}+\\\nonumber
    &+\alpha_s\beta^*_j \alpha^*_l\beta^*_m E_l D_j\rho_i E^\dagger_s D^\dagger_j\otimes\ket{0}\bra{1}+\\\nonumber
&+\alpha^*_s\beta_j\alpha_l\beta^*_m D_m E_s\rho_i D^\dagger_j E^\dagger_l\otimes\ket{1}\bra{0}+\\\nonumber
&+|\alpha_l|^2|\beta_j|^2 D_m E_s\rho_i E^\dagger_s D^\dagger_j\otimes\ket{1}\bra{1}).
\end{align}
For operators (or vacuum amplitude) for which the only no-zero value of off diagonal control qubit terms are the one for which $s=l$ and $j=m$, \cref{eq: 6} becomes
\begin{align}
   \label{eq: 13}
    \rho''_{SE}=&\sum_{lj}(E_l D_j\rho_i D^\dagger_j E^\dagger_l\otimes\ket{0}\bra{0}+\\\nonumber
    &+E_l D_j\rho_i E^\dagger_l D^\dagger_j\otimes\ket{0}\bra{1}+\\\nonumber
&+D_j E_l\rho_i D^\dagger_j E^\dagger_l\otimes\ket{1}\bra{0}+\\\nonumber
&+ D_j E_l\rho_i E^\dagger_l D^\dagger_j\otimes\ket{1}\bra{1}).
\end{align}
which is exactly the input-output relationship of a quantum switch. Although they can appear artificial, the conditions imposed on the vacuum amplitudes  are still physical as long as they respect the normalization conditions \cite{ChiKri-18}.

\section{Conclusion}
This work have shown a quantum walk approach to the quantum path dynamics. In particular it has been shown how, extending some concepts proper of the quantum walk (such as the coin tossing operator) in the context of quantum channel superposition, it's possible to obtain the input-output relationship of a quantum switch. In this work, the $X$ Pauli plays the role of the coin tossing operator in the quantum random walk, while the superposition play the role of the conditional shift. A two-hop walk dynamics of the informational carrier has the same effect of sending the carrier through a quantum switch. The result is always exact in the case of unitary channels, while there are some restrictive conditions in the vacuum amplitudes or in the Kraus operators in the case of general and non-unitary, quantum channels. A deep analysis of this conditions in terms of physical meaning and applications is an open question that can be addressed in a future work. Although preliminary, these results are very interesting because open the way towards an easier way to implement the quantum switch at an experimental level. This, in turn, opens to the possibility of perfect quantum communication even in the presence of noisy and entanglement-breaking channels.

\bibliographystyle{IEEEtran}
\bibliography{bibliografia.bib}

\end{document}